\documentclass[showpacs,aps,twocolumn]{revtex4}
\usepackage{amssymb}

\begin{document}

\title{Improving the security of secure direct communication based on secret transmitting order of particles}
\author{ Xi-Han Li,$^{1,2}$ Fu-Guo Deng,$^{1,2,3}$\footnote{Email address: fgdeng@bnu.edu.cn}
and Hong-Yu Zhou$^{1,2,3}$ }
\address{$^1$ The  Key Laboratory of Beam Technology and Material Modification of Ministry of Education,
Beijing Normal University,
Beijing 100875, People's Republic of China\\
$^2$ Institute of Low Energy Nuclear Physics, and Department of
Material Science and Engineering, Beijing Normal University, Beijing
100875,
People's Republic of China\\
$^3$ Beijing Radiation Center, Beijing 100875, People's Republic of
China}
\date{\today }

\begin{abstract}
We analyzed the security of the secure direct communication protocol
based on secret transmitting order of particles recently proposed by
Zhu, Xia, Fan, and Zhang [Phys. Rev. A \textbf{73}, 022338 (2006)],
and found that this scheme is insecure if an eavesdropper, say Eve,
wants to steal the secret message with Trojan horse attack
strategies. The vital loophole in this scheme is that the two
authorized users check the security of their quantum channel only
once. Eve can insert another spy photon, an invisible photon or a
delay one in each photon which the sender Alice sends to the
receiver Bob, and capture the spy photon when it returns from Bob to
Alice. After the authorized users check the security, Eve can obtain
the secret message according to the information about the
transmitting order published by Bob. Finally, we present a possible
improvement of this protocol.

\end{abstract}

 \maketitle


\section{Introduction}
Since an original quantum key distribution (QKD) scheme was proposed
by Bennett and Brassard \cite{bb84} in 1984 (BB84), quantum
communication has progressed quickly. There are several remarkable
branches of quantum communication, such as QKD
\cite{ekert91,bbm92,b92,gisin,longqkd,CORE,BidQKD}, quantum secret
sharing \cite{QSS1,QSS2,QSS3}, quantum secure direct communication
(QSDC), and so on. QKD whose task is to create a private key between
two remote authorized users is one of the most important
applications of quantum mechanics in the field of information. By
far, there has been a lot of attention focused on QKD
\cite{ekert91,bbm92,b92,gisin,longqkd,CORE,BidQKD}.

QSDC is a new branch of quantum communication and is used to
transmit a secret message directly without creating a private key in
advance \cite{two-step,QOTP,Wangc,Wangc2,bf}.  In 2002, Bostr\"{o}m
and Felbinger proposed a quasi-secure quantum direct communication
protocol, called "ping-pong" protocol \cite{bf}. They used
Einstein-Podolsky-Rosen (EPR) pairs as quantum information carriers
(QIC), following some ideas in quantum dense coding \cite{dense}.
However, it has been proved insecure in a noise channel
\cite{wojcik}. In 2003, Deng \emph{et al.} put forward a two-step
QSDC protocol using a block of EPR pairs \cite{two-step} and another
one with a sequence of single photons \cite{QOTP}. Wang \emph{et
al.} \cite{Wangc} introduced a high-dimension QSDC scheme.

Another class of quantum communication has been called deterministic
secure quantum communication (DSQC) \cite{lixhk} in which the
receiver can read out the secret message only after the transmission
of an additional classical bit for each qubit, different from QSDC
in which the secret message can be read out directly without
exchanging classical information anymore. Compared with QKD, DSQC
can be used to obtain a deterministic information, other than a
random binary string. Recently, Gao \emph{et al.} \cite{Gao,yan} and
Man \emph{et al.} \cite{zhangzj} proposed several DSQC protocols
based on quantum teleportation \cite{teleportation} and entanglement
swapping \cite{entanglementswapping}. Although the users have to
exchange a lot of classical information to obtain the secret
message, they can check the eavesdropping before they transmit the
secret message, and the qubits which carry the secret message need
not be transmitted again after the users check eavesdropping.
Therefore these schemes may be more secure in a noise channel and
more convenient for quantum error correction \cite{lixhk}.

Recently, Zhu \emph{et al.} \cite{zhu} proposed a new secure direct
communication protocol using EPR pairs as QIC (We called it ZXFZ
protocol for short below), similar to the two-step QSDC protocol
\cite{two-step}. The transmitting order of particles is secret to
any other people except for the sender, and the most important
advantage emphasized is that this protocol only needs one security
checking process. However, we found this scheme is insecure just due
to lack of sufficient security-checking processes. We can use the
Trojan horse attack strategy \cite{gisin} to get the secret message
completely without leaving a trace. In this paper, we first review
the protocol they proposed and then introduce the way to eavesdrop
it freely. Finally, we present a possible improvement of this secure
direct communication scheme.

\section{eavesdropping on the secure direct communication protocol}

It is well known that a crucial issue of secret communication is its
security. The security of quantum communication is guaranteed by the
principles in quantum mechanics against an eavesdropper with
unlimited powers, whose technology is confined only by the laws of
quantum mechanics. For QSDC or DSQC protocols, their security is
more important than that in QKD protocols because they are used to
transmit a secret message, other than a private key.

Now, let us start with the brief description of the ZXFZ protocol
\cite{zhu}. First, Alice and Bob agree
 that the four unitary operations $U_0=\vert 0 \rangle \langle 0
\vert + \vert 1 \rangle \langle 1 \vert$, $U_1=\vert 0 \rangle
\langle 0 \vert - \vert 1 \rangle \langle 1 \vert$, $U_2=\vert 0
\rangle \langle 1 \vert + \vert 1 \rangle \langle 0 \vert$, and
$U_3=\vert 0 \rangle \langle 1 \vert - \vert 1 \rangle \langle 0
\vert$ represent two bits of classical information 00, 11, 10, and
10, respectively. Alice prepares a sequence of EPR pairs in one of
the four Bell states, say $\vert \Psi \rangle_i
=\frac{1}{\sqrt{2}}(\vert 0 \rangle_{H_i} \vert 1
\rangle_{T_i}-\vert 1 \rangle_{H_i} \vert 0 \rangle_{T_i})$, and
then divides them into two parter-photon sequences. She keeps one
sequence (home sequence) in her laboratory and sends the other
sequence (travel sequence) to Bob through a quantum channel. After
receiving the travel sequence, Bob chooses a sufficiently large
subset of photons as a checking set (C set) and the rest as a
message set (M set). Bob encodes his checking message and secret
message by performing the four unitary operations $U_i$
$(i=0,1,2,3)$ on the C set and the M set respectively. Then Bob
rearranges the order of the T sequence and returns them to Alice.
After Alice claims her receipt of all the T sequence, Bob announces
the position of the C set and the secret order in it. Alice performs
the Bell-state measurements on the checking photons and publishes
the results. Bob can distinguish whether there is an eavesdropper
monitoring their quantum line by comparing his checking message with
Alice's outcomes. If there exists an eavesdropper, Bob terminates
the communication. Otherwise, he exposes the secret order of the M
set, and then Alice can obtain the secret message with Bell-state
measurements.

The security of the ZXFZ scheme \cite{zhu} is based on the secret
order of the particles. However, the secret order will be published
by Bob after the security checking. One can see that the two
authorized users only check the security once in the line from Bob
to Alice. The secret message is encoded with the unitary operations
done by Bob. If Alice and Bob cannot detect the eavesdropper during
the checking process, Eve can get the secret order and the whole
secret message. The eavesdropper can utilize the loophole that the
users do not check the security of the quantum channel from Alice to
Bob to insert some additive photons in each legitimate one to get
Bob's operation information freely. There are two kinds of Trojan
horse attack strategies. One is the invisible photon eavesdropping
(IPE) scheme proposed by Cai \cite{cai} and the other is the
delay-photon Trojan horse attack \cite{gisin,deng}.

Firstly, the invisible photon eavesdropping scheme utilizes the fact
that the single photon detector is only sensitive to the photons
with a special wavelength \cite{cai}. Therefore, Eve can select a
wavelength far away from that the authorized users use, which is
invisible to Bob's detector. But there exist some problems if Eve
uses the IPE to attack the quantum communication protocols in which
the wavelength-dependent optical devices are used to code the useful
information. That is, Eve maybe obtain nothing about the information
of the operations done by the legitimate users with optical devices
(such as $\lambda/2$ and $\lambda/4$ plates) if the wavelength of
the invisible photon is far away from that used by the users.
However, it is worthy to point out that no security checking is
performed in the line from Alice to Bob, which is a serious security
loophole of the ZXFZ protocol \cite{zhu}. Eve can choose a special
wavelength which is close to the legitimate wavelength to produce
the invisible photons. As we assumed \cite{gisin,wojcik}, Eve has
absolutely no technological limits for her eavesdropping; i.e., she
can do everything that quantum mechanics does not explicitly forbid.
Since the number of photons and the polarization of a photon are
commutative, Eve can insert a spy photon in each signal pulse and
sort it out without disturbing the state of the travel photon of
Alice's in principle. Now we analyze the attack scheme in detail.
First, the eavesdropper Eve prepares a sequence of EPR pairs with
the wavelength $\lambda'$ (The legitimate wavelength is $\lambda$,
$\lambda' \thickapprox \lambda$. Eve can distinguish them in
principle even though there may be no those devices existing at
present.) also in the state $\vert \Psi^{\prime}
\rangle_{i^{\prime}} =\frac{1}{\sqrt{2}}(\vert 0
\rangle_{H_{i^{\prime}}} \vert 1 \rangle_{T_{i^{\prime}}}-\vert 1
\rangle_{H_{i^{\prime}}} \vert 0 \rangle_{T_{i^{\prime}}})$. When
Alice sends the T sequence to Bob, Eve adds her $T^{\prime}$
sequence to the T sequence and forwards them to Bob. In detail, Eve
inserts each photon $T^{\prime}$ into the photon T's pulse. When Bob
performs his unitary operation on the T sequence, he also performs
his operation on the $T^{\prime}$ sequence Eve sent. After Bob
rearranges the order of the T sequence, Eve captures her spy photons
when they run back from Bob to Alice, and stores them. It is
important to point out that all of the operations Eve does have no
effect on the secret order and the secret states of the Alice's T
sequence. Since the optical devices used to accomplish the unitary
operations are often wavelength-dependent in practical, the
information carried by Eve's additional sequence $T^{\prime}$ is not
as exactly same as the photon sequence T in the line from Bob to
Alice. However, since $\lambda'$ is close to $\lambda$, the
probability that Eve obtain the correct outcome with her Bell-state
measurements is close to 1. In other words, almost all the
information about the secret message will be leaked to Eve without
being detected. After Alice and Bob accomplish the security
checking, Eve can rearrange the sequence order according to the
information published by Bob, and do the same measurements as Alice
to obtain Bob's secret message with a large probability in
principle.

Secondly, the delay-photon Trojan horse attack \cite{deng} is
inserting a spy photon in a legitimate signal with a delay time,
shorter than the time windows. As we know, in experiment there is a
"door" (a time window) of the optical device which is open only
during a short time, i.e., only when the qubits arrive. In order to
limit the Trojan horse attack, the door should be open only during a
time as short as possible \cite{trojan}. However, in practice,
timing has a finite accuracy, the eavesdropper Eve with a infinite
power can add her probes before or after the legitimate pulses.
Different from the IPE attack, the delay spy photon has the same
wavelength as the legitimate photon. Therefore the spy photon
sequence $T^{\prime \prime}$ will carry the same information as the
legitimate T sequence. Eve can prepare the  EPR pair sequence in the
same state $\vert \Psi^{\prime\prime} \rangle_{i^{\prime\prime}}
=\frac{1}{\sqrt{2}}(\vert 0 \rangle_{H_{i^{\prime\prime}}} \vert 1
\rangle_{T_{i^{\prime\prime}}}-\vert 1
\rangle_{H_{i^{\prime\prime}}} \vert 0
\rangle_{T_{i^{\prime\prime}}})$, and insert her $T^{\prime \prime}$
sequence into the T sequence when Alice sends it to Bob. In detail,
Eve inserts each $T^{\prime \prime}$ photon after each T pulse with
a delay time which is shorter than the  time windows of Bob's
optical devices. Since the $T^{\prime \prime}$ photons have the same
wavelength as the T photons, Bob will perform the exact same
operation on the $T^{\prime \prime}$ sequence when he performs his
unitary operations on the T sequence. Eve sorts out her $T^{\prime
\prime}$ photons when Bob returns them back to Alice, and rearranges
the order according to the information published by Bob after Alice
and Bob complete their eavesdropping checking. Thus Eve can perform
the Bell-state measurement on her spy photons and get the secret
message fully and freely.

Certainly, these attack schemes also work for the other QSDC
protocols, such as those in Refs. \cite{QOTP,Wangc,bf}. However, the
user can exploit a complex eavesdropping-checking process to avoid
it \cite{deng}. As there is not eavesdropping checking when the
photons are transmitted from the sender Alice to the receiver Bob in
the ZXFZ protocol \cite{zhu}, this attack cannot be detected in
principle.

\section{improvement to defeat the Trojan horse attack}

In order to defeat Eve's IPE attack, a filter with which only the
wavelengths close to the operating one can be let in should be added
before all of the Bob's devices. In this way, Eve's invisible
photons will be filtered out. Moreover, a photon number splitter
(PNS: 50/50), which is used to divide each signal into two pieces,
should be introduced to defeat the delay-photon Trojan horse attack.
Thus, with the PNS and two single-photon measurements the users can
distinguish whether there exists a multiphoton signal (including the
delay-photon signal and the invisible photon whose wavelength is so
close to the legitimate one that it cannot be filtered out with the
filter). Although  a PNS is not feasible with current technology,
the users can use a photon beam splitter (PBS: 50/50) to prevent Eve
from stealing the secret message with a little modification
\cite{deng}.

In order to improve the security of the ZXFZ protocol \cite{zhu}, we
have to take these two kinds of attacks into account. For integrity,
we describe the improved ZXFZ protocol in steps as follows.

(1) Alice prepares a sequence of EPR pairs in the state $\vert \Psi
\rangle_i =\frac{1}{\sqrt{2}}(\vert 0 \rangle_{H_i} \vert 1
\rangle_{T_i}-\vert 1 \rangle_{H_i} \vert 0 \rangle_{T_i})$ and
divides them into two sequences, the home (H) sequence and the
travel (T) sequence, same as Refs. \cite{Wangc,zhu}. She sends the T
sequence to Bob.

(2) Bob inserts a filter in front of his devices to filter out the
photon signal with an illegitimate wavelength, and then chooses a
sufficiently large subset of photons randomly. He splits each
sampling signal with a PNS and measures the two signals after the
PNS with the two measure bases $\sigma_z$ and $\sigma_x$ chosen
randomly. If the multiphoton rate is unreasonable high, Bob
terminates the transmission and repeats the communication from the
beginning. Otherwise, he continues to the next step.

(3) Bob chooses a large subset of photons from the  photons remained
as checking set (C set) and the others as the message set (M set).
He encodes his message (checking message and secret message) by
performing one of the four unitary operations $U_i$ $(i=0,1,2,3)$.
Then he disturbs the initial order of the photons in the T sequence
and returns them to Alice.

(4) After Alice announces the receipt of all the T photons, Bob
tells her the position and the order of the C set. Alice performs
the Bell-state measurement on the photons in the C set and publishes
the results with which Bob can estimates the security of the
transmission. If there is an eavesdropper, Bob stops the
communication. Otherwise, Bob publishes the order of the M set, and
Alice can obtain the secret message with Bell-state measurements.

The improved ZXFZ protocol introduces a filter and another
eavesdropping-checking process to defeat the IPE and the
delay-photon Trojan horse attacks. In principle, the eavesdropper
Eve with a infinite power can always find loopholes in quantum
communication protocols with a non-ideal quantum channel. Our
improvement can only counter the attacks we have already known. If
some sophisticate Trojan horse attacks would be put forward in the
future, we should choose a more complex eavesdropping-check way. The
sticking point we want to emphasis is that two times of security
checking is inevitable to ensure this bidirectional secure
communication, same as those in the two-step protocol
\cite{two-step}. In this way, the origin ZXFZ protocol \cite{zhu}
cannot improve the efficiency of secure communication and has no
superiority, compared with the two-step QSDC scheme \cite{two-step},
not the case announced by the authors \cite{zhu}. It is worthy to
point out that if the authorized users perform the security checking
twice, the step to disturb the order of the photons in the T
sequence can be reduced, and Alice can read out the secret message
directly without the information of the order Bob published if Bob
also has the capability of storing the quantum states. In this way,
the protocol is equivalent to the QSDC scheme proposed by Wang
\emph{et al.} \cite{Wangc}. On the other hand, the improved ZXFZ
protocol is useful if one of the two users, i.e., the sender of the
secret message (Bob), does not have the device for storing quantum
states. In this time, Bob can sample some photons from the T
sequence synchronously for checking eavesdropping in the line from
Alice to Bob, and then disturbs the orders of the others after
encoding the message. The process for eavesdropping checking of the
line from Alice to Bob needs not the storage of the other photons,
which will reduce the requirements on Bob's devices largely in a
practical application.

\section{discussion and summary}

In Ref. \cite{zhu}, the authors also proposed a one-way secure
direct communication scheme  based on the secret transmitting order
of particles with EPR pairs. We found this scheme is secure in an
ideal quantum channel. Also, the authors announced that their
one-way scheme greatly reduces the opportunity of the particles
being intercepted than the two-step protocol \cite{zhu}.
Unfortunately, we found that the opportunity of the particles being
intercepted in these two schemes is the same one. In both schemes
$2N$ ($N$ is the number of EPR pairs used) particles were
transmitted from Alice to Bob. The only difference between those two
protocols is that the $2N$ particles is transmitted in one step in
the one-way scheme \cite{zhu} and through two steps in the two-step
protocol \cite{deng}. In the two-step protocol the receiver can read
out the secret message directly. But in this one-way scheme, for
each qubit information to be understood at least one additional
classical bit of information is exchanged. Both these two protocols
need the quantum memory. Suppose the times for the transmission of
the photons  and the classical information transmitted from one user
to the other both are $t$ (Let us neglect the time for measurement
and comparison). We found that the receiver needs to store the $2N$
particles at least for $4t$ time in the one-way scheme \cite{zhu}
and he stores $N$ particles at least for $2t$ time in the two-step
scheme \cite{two-step}. In detail, in the one-way scheme \cite{zhu}
the receiver (Bob) should first tells the sender (Alice) the
information that he has received all the photons, and then Alice
publishes the positions of the photons in the checking set (i.e., C
set) and their orders. After Bob transmits the outcomes of the C set
to Alice, she tells him the orders of the other photons. That is,
Bob at least stores the $2N$ photons for four times of the time $t$.
In the two-step protocol \cite{two-step}, after receiving the
checking sequence Bob first picks out some samples and measures
them, and then he exchanges some classical information with the
sender Alice. If the transmission of the checking sequence is
secure, Alice sends the other sequence to Bob. In this time, Bob
need in principle store the checking sequence for two times of the
time $t$. From these analyses above, we can see that the two-step
protocol is more convenient than the one-way scheme proposed by Zhu
\emph{et al.} \cite{zhu}.

In summary, we analyzed the security of the secure direct
communication proposed by Zhu \emph{et al.} \cite{zhu} and found
that this protocol is insecure because the two parties only execute
one security-checking process. We proved that the eavesdropper can
get Bob's secret message with a large probability without being
detected by using the invisible photon eavesdropping scheme
\cite{cai} or get all the secret message with the delay-photon
Trojan horse attack  \cite{deng}. We also present a possible
improvement of this protocol by introducing a filter and another
complex security-checking process to defeat these two kinds of
attacks. The most important point is that for each block of
transmission, an eavesdropping checking is inevitable for secure
communication no matter what is transmitted with a quantum channel.

\section*{ACKNOWLEDGMENTS}

This work is supported by the National Natural Science Foundation of
China under Grant No. 10604008, 10435020, 10254002 and A0325401, and
Beijing Education Committee under Grant No. XK100270454.

\end{document}